\documentclass[article]{JHEP3}
\usepackage{epsfig}

\title{Numerical Study of  Cosmic Censorship
in String Theory}
\preprint{\hepth{0402109}}
\author{M. Gutperle\\
Department of Physics and Astronomy, UCLA, Los Angeles, CA 90095, USA
\\{\tt E-mail:
gutperle@physics.ucla.edu}}
\author{P. Kraus\\Department of Physics and Astronomy, UCLA, Los Angeles, CA 90095, USA
\\{\tt E-mail:
pkraus@physics.ucla.edu}}
\abstract{  Recently Hertog, Horowitz, and Maeda have argued that
cosmic censorship can be generically violated in string theory in anti-de
Sitter spacetime
by considering a collapsing bubble of a scalar field whose mass saturates
the Breitenlohner-Freedman bound.   We study this system numerically and  find that for various choices of initial data  black holes form rather than naked
singularities, implying that in these cases  cosmic censorship is upheld.

}

\def\bea{\begin{eqnarray}}
\def\eea{\end{eqnarray}}
\def\be{\begin{equation}}
\def\ee{\end{equation}}
\def\half{{1\over 2}}

\def\tx{\tilde{x}}
\def\hx{\hat{x}}

\begin{document}

\maketitle
\section{Introduction}

Inasmuch as they mark a breakdown of classical general relativity,
spacetime singularities pose an important challenge to theories of
quantum gravity such as string theory.  Singularities are common
occurrences in the study of cosmology and black holes, but their correct
physical description still seems to be beyond our grasp. Penrose
proposed the concept of cosmic censorship, which states that
singularities which are formed by nonsingular initial data are always
hidden behind a horizon \cite{PenrosePC}.

 In the case of
cosmology it has proven quite challenging to find controlled
examples within string theory
\cite{Balasubramanian:2002ry}\cite{LiuFT}\cite{LiuKB}\cite{Craps:2002ii}\cite{Fabinger:2002kr}\cite{CornalbaFI}\cite{CornalbaNV}
\cite{GutperleAI}, while for black holes the existence of the
event horizon makes it difficult to study the physics of the
singularity itself
\cite{Maldacena:2001kr}\cite{KrausIV}\cite{FidkowskiNF}\cite{Horowitz:2003he}\cite{Gottesman:2003up}.

To make progress it would be very helpful to find a naked singularity
within a well understood background of string theory.
Finding a naked singularity in an anti-de Sitter background would be
especially promising, as the nonperturbative tools of the AdS/CFT
correspondence could be brought to bear on the problem.   An interesting
 proposal for obtaining such singularities was put forward in a
recent paper by Hertog, Horowitz and Maeda (HHM) \cite{HertogXG}(
see also \cite{HertogZS}).  Their argument uses the special
properties of scalar fields which saturate the
Breitenlohner-Freedman bound
\cite{BreitenlohnerJF}\cite{BreitenlohnerBM}, which describes how
tachyonic a field in AdS can be without leading to an instability.
If one starts with a large and homogeneous bubble of such a scalar
field the resulting evolution leads to a singularity in finite
time.  To argue that the singularity is naked, HHM claim that the
energetics of the system is such that no black holes can form.

In this paper we present results on the numerical study of such
collapsing bubbles.  For all the cases we have studied, we will
see that the evolution is consistent with the formation of black holes
rather than naked singularities. HHM have actually
considered two distinct scenarios.  For the first, one imposes
standard boundary conditions in AdS, and the argument revolves
around the fact that the initial data has negative mass; black holes
on the other hand have positive mass, which might seem to
prevent them from forming from the assumed initial data.  However,
there is a loophole: this definition of the mass is not
conserved, as HHM themselves point out.
   Numerically we will find
that the mass rapidly becomes positive, and that a trapped surface
subsequently forms outside the bubble.
 In the second scenario HHM impose Dirichlet boundary
conditions at some cutoff position inside AdS.  In this case, the mass
is conserved, but one also has to consider the possible existence
 of hairy black  holes whose mass is lower than that of the initial data
(the no hair theorem does not apply in the presence of the
cutoff).  We will compare the mass of the black hole with the mass
of the initial data for various choices of parameters and find
that  black holes have sufficiently small mass to form. Similarly,
our numerical investigations of the collapsing bubble with
Dirichlet boundary conditions again leads to the conclusion that
black holes form rather than naked singularities.  Thus, for all
the examples we have studied cosmic censorship is upheld. However,
we should note that we have by no means exhausted all possible
choices of parameters, including some suggested in
\cite{HertogXG}, and so it is possible that cosmic censorship can
be violated in some other regime.

\section{Equations of motion}

We consider the action for a scalar field coupled to gravity in 5 dimensions with a negative cosmological constant,
\be
S =\int \! d^5 x \, \sqrt{-g} \Big(\half R+6  - \half (\partial
\phi)^2 -V(\phi) \Big).\label{aaa}
\ee
The potential corresponding to a free scalar which saturates the
Breitenlohner-Freedman bound is
\be
 V(\phi) =-2\phi^2. \label{aba}
\ee
For now we will assume \ref{aba}, although we will also consider
the full potential arising in supergravity in section 3.  We have
chosen our units such that the background AdS solution is
\be
 ds^2=-(1+r^2)dt^2 + {dr^2 \over 1+r^2}  +r^2 d\Omega_3^2,
\quad \phi=0.\label{ax}
\ee

Spherical symmetric solutions will be considered throughout.
The general such metric in the ADM decomposition is
\be
ds^2= - N^t(t,r)^2 dt^2 +L(t,r)^2(dr+N^r(t,r) dt)^2 +R(t,r)^2 d\Omega_3^2.\label{ab}
\ee
It is very convenient to choose $R(t,r)=r$ so that the coordinate $r$ has a clear physical meaning.  In our numerical analysis we
will be looking for the appearance of trapped surfaces, and we now state the condition for their existence.  Consider radial and
future  directed null geodesics, obeying
\be
{dr \over d\lambda} = \Big(-N^r \pm {N^t \over L}\Big){dt \over d\lambda}.\label{ac}
\ee
A trapped surface exists whenever ${dr \over d\lambda} <0$ for either choice
of sign. This can be rewritten in a more useful form as follows.
Define
\be
 \mu = r^4 +r^2 - {r^2 \over L^2} +\left({rN^r \over N^t}\right)^2.\label{ad}
\ee
$\mu$ is proportional to the quasilocal mass, and essentially
measures the total amount of energy contained within an $S^3$ of
radius $r$.  If we solve (\ref{ad}) for $L$ and substitute into (\ref{ac}) we
find
\be
 {dr \over d\lambda} = \left( -N^r \pm
\sqrt{ (N^r)^2 + \left(1+r^2 -{\mu \over r^2}\right)(N^t)^2}
~\right) {dt \over d\lambda}.\label{ae} \ee
We conclude that trapped surfaces occur for $1+r^2 -{\mu \over r^2}<0$.
For given constant $\mu$, this is the same as saying that $r$ is inside
the horizon of the AdS-Schwarzschild black hole.  We will refer to
 $1+r^2 -{\mu \over r^2}<0$ as the ``horizon function''.

$\mu$ was defined above in any coordinate system of the  form  (\ref{ab}) with
$R=r$.  In our numerical study we will choose to set $N^r=0$ and
write the metric as
\be ds^2 =- n(t,r)^2 (1+r^2 - {\mu(t,r) \over r^2}) dt^2 + {dr^2
\over
 1+r^2 - {\mu(t,r) \over r^2}} + r^2 d\Omega_3^2.\label{af}
\ee
Setting $n=1$ and $\mu=$ constant
gives the standard form of the AdS-Schwarzschild
black hole, the mass being $M=  3\pi^2 \mu$.
Since the definition of $\mu$ in (\ref{af}) agrees with that in
(\ref{ad}), we again see
that trapped surfaces occur for  $1+r^2 -{\mu \over r^2}<0$.  But given
the form of the metric (\ref{af}), it is clear that the coordinates will break
down precisely when trapped surfaces are encountered.  Therefore in our
numerics we will look for $1+r^2 -{\mu \over r^2}$ becoming very small,
and assume that the subsequent evolution is sufficiently smooth that
trapped surfaces do indeed occur. We will also give evidence that the trapped
surfaces are associated with black hole formation, although it would
be interesting to check this directly by using coordinates which
can extend into the region containing the trapped surfaces.

With spherical symmetry, the gravitational field has no dynamical
degrees of freedom, and the metric functions $n$ and $\mu$ on a given
time slice can be determined in terms of $\phi$ and $\dot{\phi}$
evaluated on the same time slice.  In particular, two linear combinations
of Einstein's equations yield the constraints
\bea
\mu'&=&-{4\over 3}r^3 \phi^2+ {1\over 3}(1+r^2
-{\mu\over r^2}) \Big( {\pi_\phi^2 \over r^3}+ r^3(\phi')^2 \Big),
\label{aga}\\
{n'\over n }&=&  {1\over 3 r^2 } \Big( {\pi_\phi^2 \over r^3}+ r^3(\phi')^2
\Big).\label{ag}
\eea
Here, $\pi_\phi$ is the canonical momentum
\be
\pi_\phi = {r^3 \over n (1+r^2 - {\mu(t,r) \over r^2})} \dot{\phi}.\label{az}
\ee
We also need the scalar field equation of motion, which is
\bea
\dot \pi_\phi&=& \Big( r^3 (1+r^2 -{\mu\over r^2})
n \phi'\Big)'+  4 r^3
n \phi\nonumber\\
&=&  n\Big(r^3 (1+r^2-{\mu\over r^2}) \phi'' + ( 3r^2 + 5r^4 -\mu +{4\over
3} r^4 \phi^2 )\phi' + 4 r^3 \phi\Big)  \label{ah}
\eea
where (\ref{aga}) and (\ref{ag}) were used.

For numerical purposes it is useful to redefine variables.  We map
 $r\in[0,\infty]$ to
$x\in[-1,1]$ by
\be
r^2={1+x\over 1-x}.\label{ai}
\ee
The asymptotic behavior of the scalar field can be simplified by defining
\bea
 \psi &=& (1+r^2)\phi
= {2 \over 1-x} \phi \\
 \pi_\psi
&=& (1-x)^{-1/2}
(1+x)^{-3/2} \pi_\phi. \label{aj}
\eea

We now discuss the boundary conditions.  In solving (\ref{aga}) and (\ref{ag}) we can
choose the
values of $\mu$ and $n$ at the origin, and we choose these to be
\be
  \mu(t,r=0)=\mu(t,x=-1)=0, \quad  n(t,r=0)=n(t,x=-1)=1.\label{ak}
\ee
Note then that our time coordinate measures the proper time at the origin.
As for the scalar field at the origin,
we need only demand that $\psi$ is nonsingular there.   We also need
a boundary condition on the scalar field at infinity.  Solving the
free scalar field equation in AdS yields the asymptotic behavior
\be
\phi \sim {\alpha(t) \over r^2} + \beta(t) {\ln r \over r^2} \quad
\Rightarrow \quad \psi(x) \sim \tilde{\alpha}(t) + \tilde{\beta}(t)
\ln (1-x).\label{al}
\ee
The two asymptotic behaviors correspond to the normalizable and
non-normalizable modes familiar in the study of fields in AdS. We
are allowed to freely specify $\beta(t)$ as well as the values of
$\phi$ and $\dot{\phi}$ on an initial time slice, and then
$\alpha(t)$ is determined by the equations of motion (and
analogously for $\psi$, $\tilde{\alpha}$ and $\tilde{\beta}$).
Fluctuations around the ordinary AdS vacuum correspond to setting
$\beta(t)=\tilde{\beta}(t)=0$; on the CFT side of the AdS/CFT
correspondence this corresponds to evolution with respect to the
unperturbed CFT Hamiltonian.  We will refer to these boundary
conditions as the ``standard'' ones.    Besides the standard
boundary conditions, HHM consider a second set of Dirichlet
boundary conditions, which we'll return to in the next section.
So, expressed in terms of $\psi$, the standard boundary conditions
just say that $\psi$ should be finite as $x\rightarrow 1$, but is
otherwise unrestricted.

The equations of motion can now be written
\bea
\dot \psi &=&2 \Big(
2 -{(1-x)^2\over 1+x} \mu\Big) n \pi_\psi, \label{solga} \\
\dot\pi_\psi&=& {1\over 2}(1-x)  \Big( 2(1+x)-(1-x)^2 \mu\Big) n
\psi''-\Big(1 +{(1-x)x\over 1+x} \mu\Big)n
\psi\nonumber\\
& +& n  \Big((1-3x) + {(1-x)^2 (1+2x)\over 1+x}\mu +{1\over 6}
(1-x)^2(1+x) \psi^2  \Big)\psi'\nonumber\\
&-&{1\over 6} (1-x) (1+x) n \psi^3, \label{solg}
\eea
where the metric function $n$ can be expressed as follows
\be n= e^{  \int_{-1}^x \! d\tx \,f(\tx)} \ee and $\mu$ is given
by \bea \mu &=&e^{-\int_{-1}^x \! d\tx \,f(\tx)}\int_{-1}^x \!d
\hx \Big\{-{1\over 6}{(1+\hx)}\psi^2 + {2\over 3}
{(1+\hx)}\pi_\psi^2
\nonumber \\
 & & \quad\quad \quad+{1\over
6 } (1+\hx)^2 (1-\hx)(\psi')^2 -{1\over 3} (1+\hx)^2 \psi'\psi \Big\}
 e^{  \int_{-1}^{\hx} \! d\tx \,f(\tx)} \label{cona}
\eea
and we defined
\be f = {1\over 3} \Big(
{(1-x)^2} \pi_\psi^2 + {1\over 4} (1- x^2) ( (1-x)
\psi' -\psi)^2\Big).\label{ua}
\ee
Note that the only potentially singular terms are of the form
${\mu \over 1+x}$, but $\mu \sim (1+x)^2$ as $x \rightarrow -1$, so this is not
a problem.

Given the way in which $\mu$ appears in the metric, its value at infinity
gives a natural definition (up to a factor of $3\pi^2$)
 of the total mass associated with the spacetime.
However, for scalar fields which saturate the Breitenlohner-Freedman
bound, it turn out that this definition of mass is not conserved.  From
the definition (\ref{cona}) and the equations of motion, one can show
\be
 {d \over dt} \mu(t,x=1) = -{2 \over 3} {d \over dt}
\left[ \psi(t,x=1)^2\right].\label{am}
\ee
A conserved mass can therefore be defined as
\be
 m = \mu(t,x=1) + {2 \over 3}  \psi(t,x=1)^2.\label{an}
\ee
Another important fact is that $\mu$ can be negative, while $m$ obeys
a positivity theorem.

Following HHM,  our initial data  will consist of a large homogeneous
bubble of scalar field, with vanishing time derivative, and with the
standard AdS boundary condition at infinity.  So at $t=0$ we have
\bea
 \psi &=& {2 \over 1-x} \phi_0   \quad {\rm for}
\quad x<x_0 \\
  \psi & \sim&  ~ {\rm finite} \quad  \quad{\rm for} \quad x
\rightarrow 1 \\
 \pi_\psi& =&0 \label{ao}
\eea
with $\phi_0$ being the constant value of $\phi$ inside the
bubble. We take the bubble to be sufficiently large that a
singularity is guaranteed to form at the center of the bubble
within finite time.\footnote{See  \cite{HertogXG} for a discussion of the
required size.} Furthermore, we restrict the size of the bubble
and the scalar field profile outside the bubble to be such that
the asymptotic value of $\mu$ is negative.  Were $\mu$  conserved
this initial data would certainly evolve to a naked singularity,
since black holes have $\mu>0$.  However, we know that $\mu$ is
not conserved, and the question becomes whether it can increase
sufficiently rapidly to allow a black hole to form before a naked
singularity appears; we will see that this is indeed what happens.

\section{Results}

We solve the equations (\ref{solga}) and (\ref{solg})  numerically
 using  C code
 by evaluating the
integrals for $n$ and $\mu$ on a given time slice, and then
evolving $\psi$ and $\pi_\psi$ forward by a simple time stepping.
We take the number of spatial lattice points to be a few thousand.
Stability requires that the time step is approximately the square
of the spatial lattice spacing, and we have taken $dt = dx^2/2$.
We have done the
 numerical evolution using various number of lattice points, and a
 refinement of the lattice does not show any sign of a numerical
 instability. In the course of the evolution $n(x,t)$ becomes very
 large near the boundary. In order to avoid a numerical instability we
 have to adjust the time stepping $dt\to dt/n(t,x)$.  We evolve in
 time until the numerics break down, which occurs when the horizon
 function $1+r^2- {\mu \over r^2}$ becomes very small, and this is
 associated with the appearance of trapped surfaces.

\subsection{Standard boundary conditions}
With $4001$ lattice points, we have chosen various values for
$\phi_0$ and $x_0$ in the initial data.  Outside the bubble, the
initial data for $\psi$ is taken to be linear in $x$, with slope
matched to the first derivative of $\psi$ just inside the bubble.
In the following we will present the results for two cases. In the
first we present a fairly generic example using the potential
\ref{aba}, and in the second we instead use the full supergravity
potential.

\medskip

\noindent{\bf Case 1.  Quadratic potential:} For $x_0=.583$ and
$\phi_0= 0.2$, we find that our numerics break down at
$t_{final}=0.8837$. The initial profile of $\psi(x)$ is given by
Fig. 1 \FIGURE[h]{ \epsfig{file=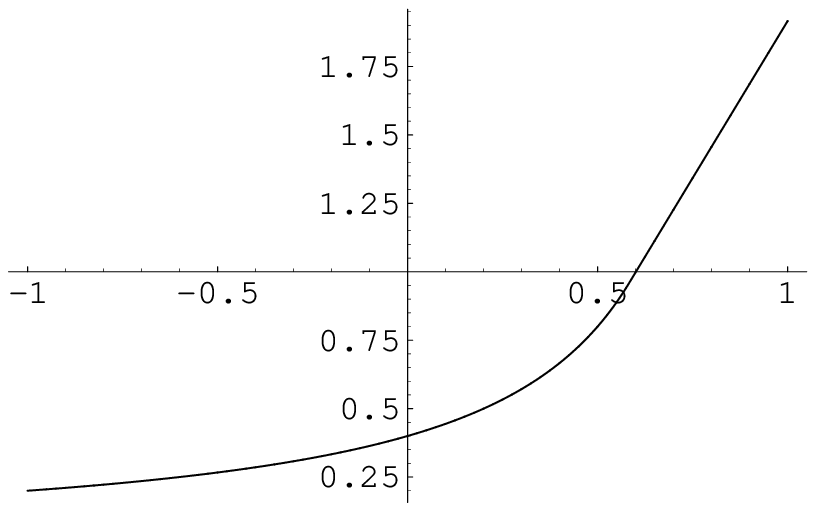,height=6cm}
\caption{Initial profile of $\psi(x)$ at $t=0$} }
\pagebreak

\noindent whereas the final profile is given by Fig. 2
\FIGURE[h]{
\epsfig{file=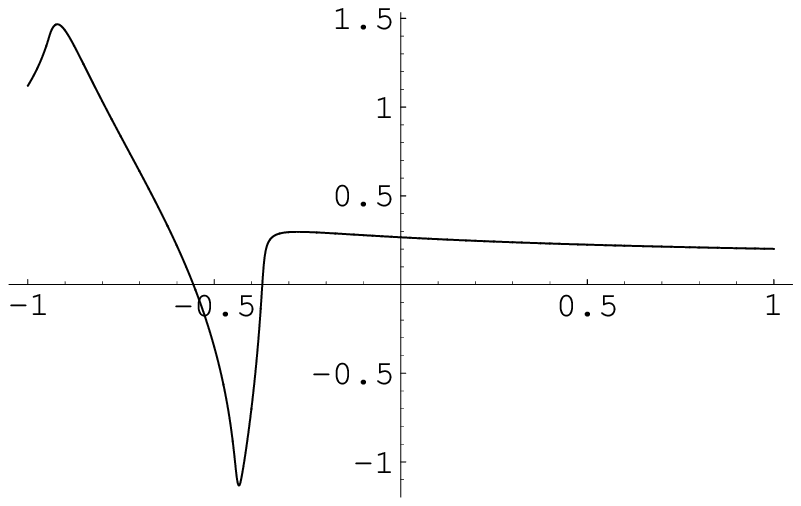,height=6cm}
\caption{Final profile of
$\psi(x)$ at $t=t_{final}$}
}

\medskip

\noindent The plot of the conserved mass $m(t)$ versus nonconserved mass
$\mu(t)$ in Fig. 3 reveals that indeed $m(t)$ is conserved all the
way to $t_{final}$. The fact that  $m(t)$ is conserved to great
accuracy all the way to $t=t_{final}$ constitutes a good check of
our numerics.
\FIGURE[h]{
\epsfig{file=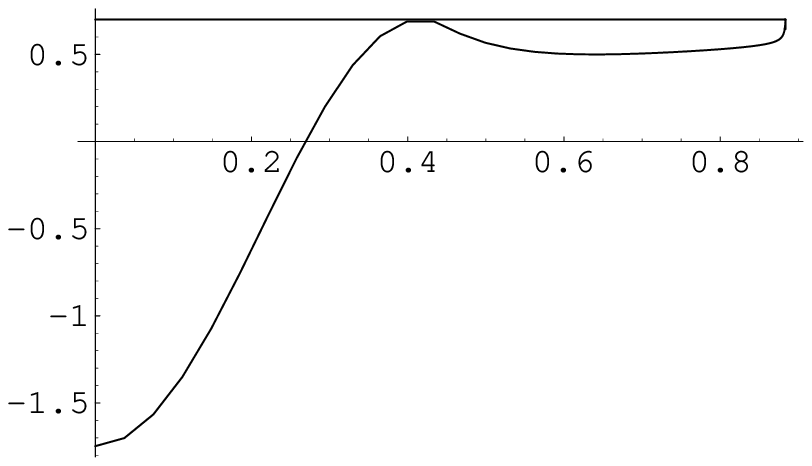,height=6cm}
\caption{Conserved mass versus nonconserved mass as functions of time}
}

\pagebreak

\noindent As  discussed in section 2, the vanishing of the horizon function
$1+r^2-\mu/r^2$ indicates the occurrence of a horizon (or more
accurately an apparent horizon)  at $x_h=-0.376$, as seen in Fig.
4
\FIGURE[h]{
\epsfig{file=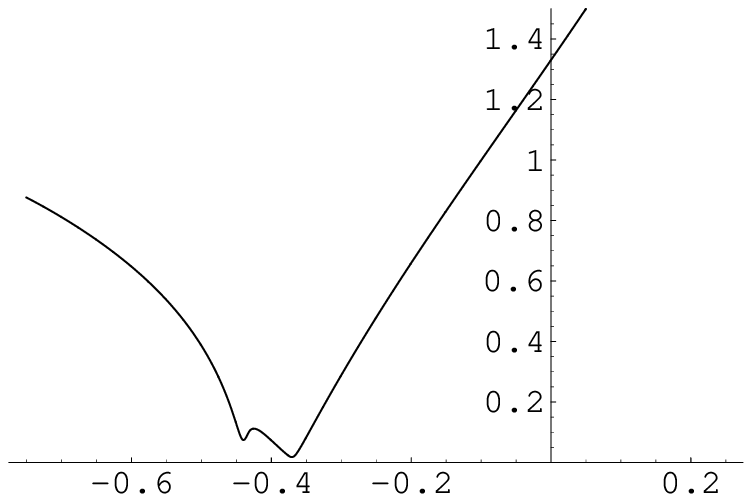, height=6cm}
\caption{Horizon function at $t=t_{final}$}
}

\medskip

\noindent The Ricciscalar remains well behaved near $x=x_h$, although it grows
in the region $x<x_h$. At $t=t_{final}$ the Ricciscalar is shown in Fig. 5
\FIGURE[h]{
\epsfig{file=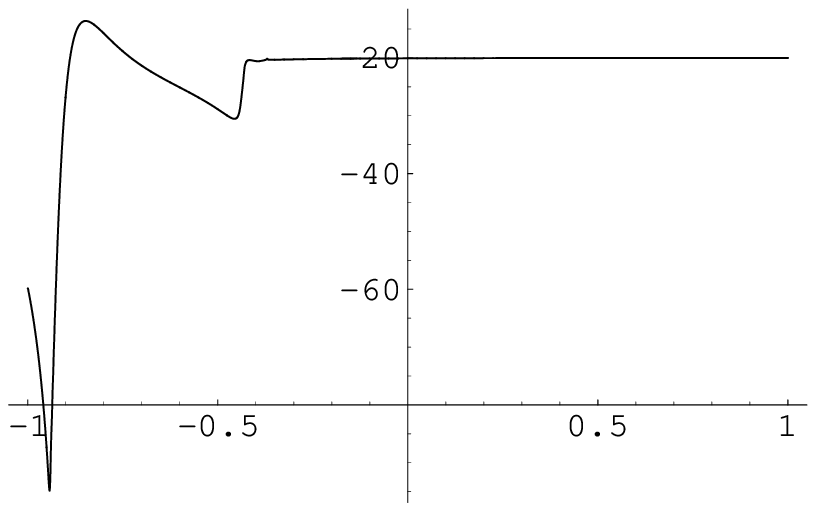, height=6cm}
\caption{Ricciscalar at $t=t_{final}$}
}

\pagebreak

\noindent {\bf Case 2.  Supergravity potential:} The quadratic
potential (\ref{aba}) for the scalar field $\phi$ is an
approximation to the exact supergravity potential \be V(\phi)= - 2
\exp({2\over \sqrt{3}}\phi)-4 \exp(-{1\over \sqrt{3}}\phi)+6 \ee
Numerically, this is more expensive to evaluate. For the same
initial data as shown in Fig.1, with $x_0= 0.583$ and $\phi_0=0.2$
and 2001 lattice points, the numerical evolution breaks down after
$t_{final}= 0.8217$. The results  are very similar to the ones for
the quadratic potential. The final profile is plotted in Fig. 6.

\FIGURE[h]{ \epsfig{file=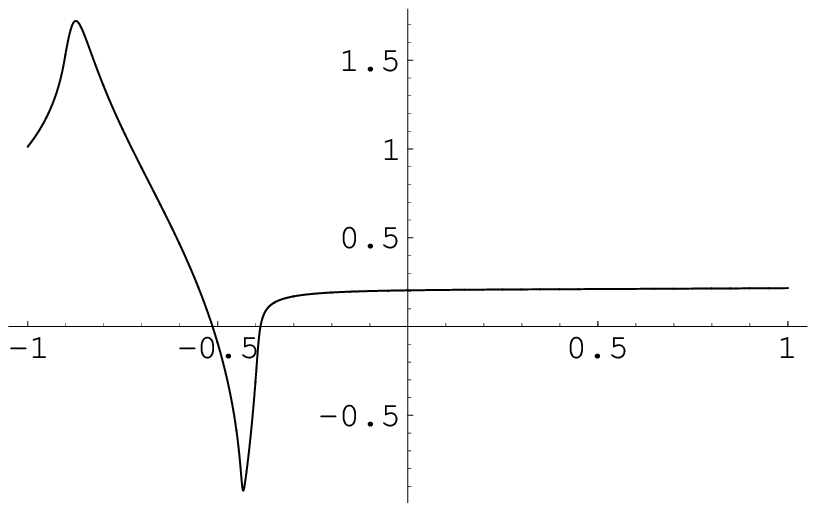,height=6cm}
\caption{Final profile of $\psi(x)$ at $t=t_{final}$} } 

\medskip

\noindent The
horizon function at $t=T_{final}$, as plotted in Fig. 7, indicates
that the horizon forms
 near $x_h=-0.41$.

\FIGURE[h]{ \epsfig{file=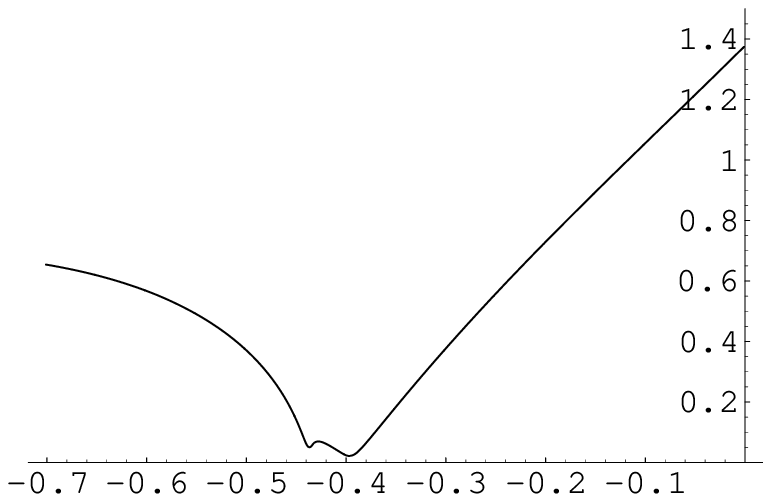, height=6cm}
\caption{Horizon function at $t=t_{final}$} } 
\noindent Hence the full
nonlinear potential displays the same characteristic features as
the quadratic potential.
\medskip

These two numerical results are a sample of the generic features
of almost all initial configurations we have studied. The
nonconserved mass  rapidly becomes positive and an apparent horizon forms.
The size of the horizon grows with the conserved mass, and we have
studied initial data where the horizon radius is considerably
larger than $1$.  We interpret the apparent horizon to be associated with
black hole formation since it appears well outside the bubble in a region
where the curvature seems to remain small.   In particular, we are 
distinguishing our trapped surface from those trapped surfaces which we know
will appear in the homogeneous region of the bubble as the singularity is
approached; the latter have nothing directly to do with black hole formation.

 We also seem to find the first
indications for the decay of the scalar field, in line with the
no-hair theorem,  and a likely endpoint of the evolution is an
AdS-Schwarzschild black hole
 without scalar hair. However, since our
numerics break down when the horizon forms, one has to use more
sophisticated numerical methods to settle this question.  In any
case, it seems clear that cosmic censorship is not violated in
this setup.

\subsection{Dirichlet boundary conditions }

 HHM have
proposed a second scenario for producing naked singularities,
which involves imposing Dirichlet boundary conditions at some
large but finite radius, $r=R_1$.\footnote{We thank Thomas Hertog
and Gary Horowitz for discussions regarding the material in this
section.}  We take constant $\phi = \phi_0$ inside a bubble of
radius $R_0$, and set $\phi(R_1)=\phi_1$.     To minimize the
energy we take $\phi(r)$ to obey the linearized equations of
motion outside the bubble:
\bea
 \phi &=&  \phi_0 \quad\quad (r<R_0) \nonumber\\
\phi &=&  {a \over r^2}+ b{\ln (1+r^2) \over r^2}
 \quad\quad (R_0<r<R_1)\label{ba}
\eea
where $a$ and $b$ are determined by the boundary conditions at
$R_0$ and $R_1$.   HHM actually consider slightly different
initial data which do not satisfy the equations of motion outside
the bubble, but we consider \ref{ba} since it will clearly give
lower mass and thus be more favorable for violating cosmic
censorship.

We will reduce to a three parameter family of initial data by
minimizing the mass with respect to $\phi_0$ while holding $R_0,
 R_1$ and $\phi_1$ fixed.  This yields the mass formula
\be \mu_{\rm initial}=-{2\over 3} \phi_1^2 R_1^2
\Bigg(1+R_1^2\Big( 1- {2+R_0^2\over (2+R_0^2)\ln(1+R_1^2)
-(2+R_0^2)\ln(1+R_0^2)+2R_0^2}\Big)\Bigg).\label{zz} \ee
Following HHM, we  further require $\phi_0 \ll 1$ so as to be able
to neglect backreaction.   In this regime, HHM show that if a
black hole forms then the size of the event horizon on the initial
surface must be at least $\phi^{2/3}_0 R_0$.

For $ R_1 \gg R_0 \gg 1$, which HHM consider, we have $\mu_{\rm
initial}<0$,  and $\mu$ is conserved due to the Dirichlet boundary
condition. But to establish a violation of cosmic censorship one
needs to further argue that any black hole obeying the same
Dirichlet condition at $R_1$, and with horizon size at least
$\phi^{2/3}_0 R_0$, has $\mu_{\rm bh}> \mu_{\rm initial}$.

We therefore look for static black hole solutions, satisfying \bea \mu' +{4\over 3} r^2 \phi^2 -{1\over 3} (1+ r^2 -{\mu\over
r^2}) r^3 (\phi')^2&=0\cr r^2(1+ r^2 -{\mu\over r^2}) \phi'' +( 3 r^2 + 5r^4 -\mu +{4\over 3} r^4
  \phi^2 )\phi' + 4 r^3 \phi&=0.\label{zw}
\eea We take the  horizon to be at $r=R_s$, and so $\mu(R_s)=
R_s^4 +R_s^2$.  The equations of motion can be seen to fix
$\phi'(R_s)$ in terms of $\phi(R_s)$, and so the black hole
solution is completely specified by $R_s$, $R_1$ and $\phi_1$. We
numerically integrate the equations outward from the horizon to
$R_1$, and then adjust $R_s$ and $\phi(R_s)$ so as to satisfy
$\phi(R_1)=\phi_1$.   We also have to satisfy $R_s> \phi_0^{2/3}
R_0$ in order to be sure that our  horizon is large enough to
enclose the region in which we know singularities will form.
Although we were able to use the free field approximation to
compute the mass of the bubble, it turns out that backreaction is
quantitatively important for the black hole, and so we are forced
to solve the full nonlinear equations \ref{zw}.

We have carried out this procedure for a few parameter choices, as
summarized below.

\vspace{.5cm} \noindent {\bf Example 1:} We take  \be R_1=20000,
\quad R_0=100, \quad \phi_1= 10^{-6}. \ee Then $\phi_0=0.006$,
hence the backreaction for the bubble can indeed be neglected. The
mass of the bubble is then $\mu_{\rm initial}= -98198.5$ and the
critical horizon radius is $\phi_0^{2/3}R_0= 3.429$.  We then
succeed in finding a black hole solution with $R_s=4$ and
$\mu_{\rm bh}=-100165$.   Since $\mu_{\rm bh} < \mu_{\rm initial}$
the energetics allow black hole formation, and so there is no
reason to expect cosmic censorship to be violated.

\vspace{.5cm} \noindent {\bf Example 2:} We take  \be R_1=100000,
\quad R_0=500, \quad \phi_1= 10^{-7} \ee yielding $\mu_{\rm
initial} = -613742$, and the critical horizon radius is
$\phi_0^{2/3}R_0= 3.699$.  We then find a black hole solution with
$ R_s=6$ and $ \mu_{\rm bh}= -630601$.   So we again find
$\mu_{\rm bh}< \mu_{\rm initial}$.

It is possible that there exists some other choice of parameters
such that $\mu_{\rm bh}> \mu_{\rm initial}$, but we have not
encountered such a case so far.

Now we turn to the numerical time evolution  in the Dirichlet
case. Using our variables, the regime $\ln R_1 / \ln R_0 \gg 1 $
considered by HHM is rather inaccessible due to the large
hierarchy of scales. We will instead consider similar initial data
as we studied in the case of standard boundary conditions, and
return to the $\ln R_1 / \ln R_0 \gg 1 $ case in a future
publication.

For the numerical evolution we chose initial data with
$x_0=0.583$, $\phi_0=0.2$,  and with 3981 lattice points. The
Dirichlet boundary condition is imposed at $x_1= 0.99$.
 The numerical evolution breaks down at $t_{final}=
0.9139$.

\pagebreak

 \noindent The initial profile of
$\psi(x)$ is shown in Fig. 8

\FIGURE[h]{
\epsfig{file=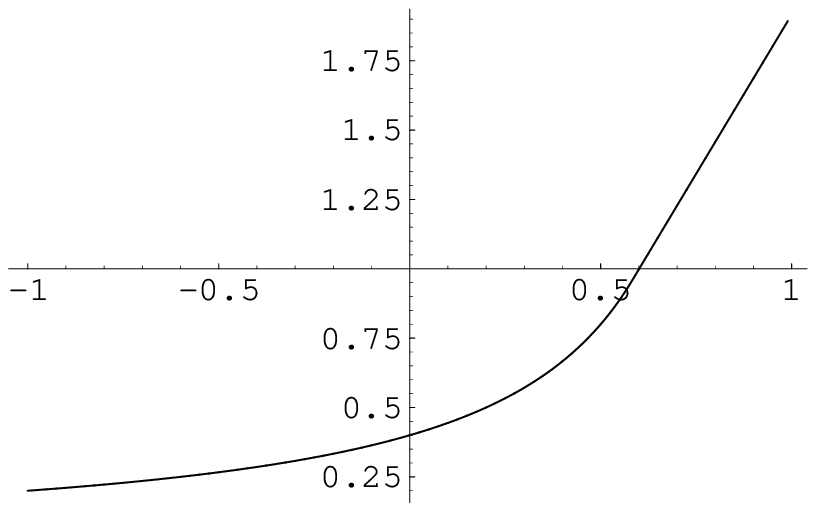,height=6cm} \caption{Initial
profile of $\psi(x)$ at $t=0$} }

\noindent and the final profile is shown in Fig. 9. 
\FIGURE[h]{
\epsfig{file=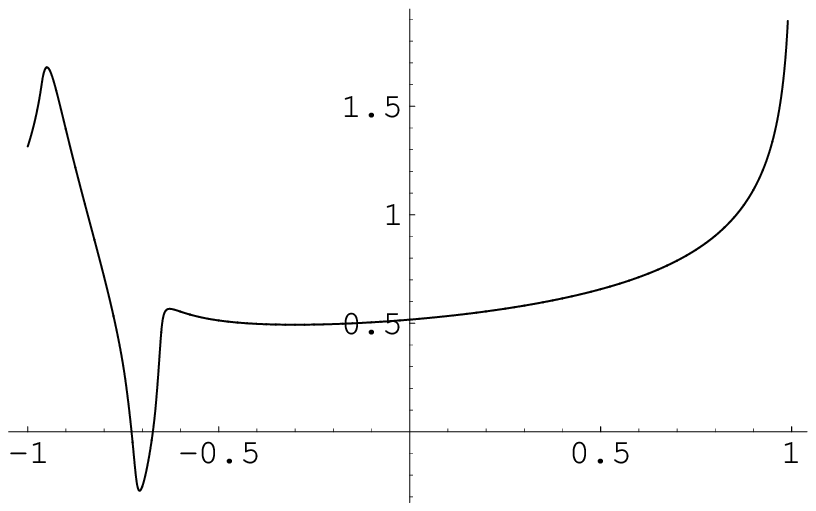,height=6cm} \caption{Final
profile of $\psi(x)$ at $t=t_{final}$} }  

\pagebreak

\noindent The horizon
function at $t=t_{final}$ is plotted in Fig. 10,
 \FIGURE[h]{
\epsfig{file=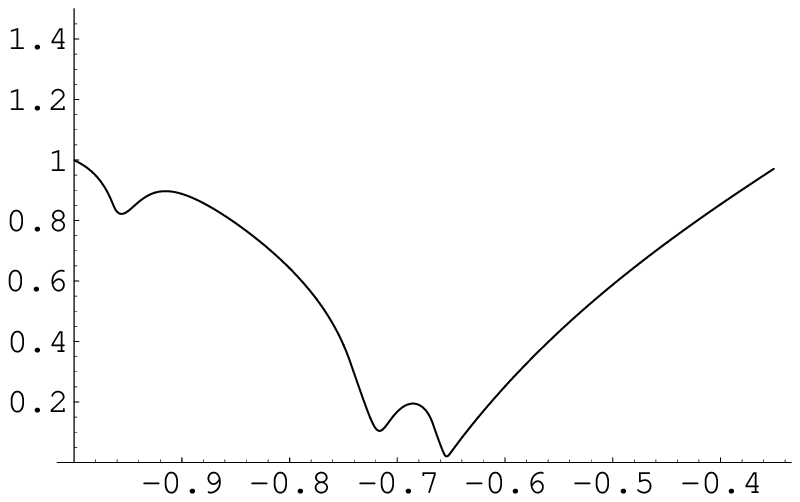, height=6cm} \caption{Horizon
function at $t=t_{final}$}}
 and the Ricciscalar is well behaved
near the horizon.   We have also checked that the mass is
conserved to excellent accuracy.  The approximate vanishing of the
horizon function  indicates that the evolution with Dirichlet
boundary conditions leads to a singularity shielded by a horizon
instead of a naked singularity.

\section{Discussion}

We have given evidence that cosmic censorship survives the
challenge from collapsing bubbles of Breitenlohner-Freedman mass
scalar fields in AdS. In particular, the existence of negative
mass for the initial data is consistent with black hole formation
since the mass becomes positive rather early in the evolution.  In
the case of Dirichlet boundary conditions there is still a
possibility of violating cosmic censorship since we have not
exhausted the full parameter space.   However, we should point out
that if one truly wishes to form a naked singularity in {\it
string theory} then we are presumably restricted to the standard
boundary conditions; being a quantum theory of gravity, string
theory only admits asymptotic boundary conditions and not
Dirichlet conditions at finite locations.

There have been a number of  previous numerical studies of the dynamics of
scalar fields coupled to gravity in AdS spacetimes
\cite{PretoriusYU}\cite{HusainNK}. In these papers the main focus was on
massless scalar fields. The study of gravitational collapse and black
hole/singularity formation using numerical techniques is a very
interesting problem in general.
 
In the future we hope to study various different scalar field
potentials, for example those considered in 
\cite{HertogZS}\cite{AlcubierreTC}, and to employ coordinates which
continue through the horizon of the black hole.    Our 
numerics were based on the simplest discretization and evolution
schemes, and it would be interesting to apply more sophisticated 
techniques like mesh refinement or excision to this problem and check
our results. In particular it would be interesting to 
investigate the critical phenomena associated with gravitational
collapse (see \cite{Gundlach:2002sx}\cite{Choptuik:xj} for 
reviews)  in the context of the AdS/CFT correspondence.

\acknowledgments

\noindent

We thank Thomas Hertog, Gary Horowitz, Veronika Hubeny, Juan Maldacena, and Steve Shenker
for discussions.
The work of MG is supported in part by NSF grant 0245096,
and the work
of PK  is supported in part by NSF grant 0099590. Any opinions,
findings and conclusions expressed in this material are those of the
authors and do not necessarily reflect the views of the National
Science Foundation.

\end{document}